\documentclass{emulateapj}
\usepackage{apjfonts}
\usepackage{amsmath}
\usepackage{graphicx}
\usepackage{algorithmic}
\usepackage{algorithm}
\shortauthors{Carter \& Agol  2012}
\shorttitle{QATS}

\begin{document}

%
\def\ltsima{$\; \buildrel < \over \sim \;$}
\def\lsim{\lower.5ex\hbox{\ltsima}}
\def\gtsima{$\; \buildrel > \over \sim \;$}
\def\gsim{\lower.5ex\hbox{\gtsima}}

\newcommand{\argmax}{\operatornamewithlimits{argmax~}}
                                                                                          
%


\title{The Quasiperiodic Automated Transit Search Algorithm}

\author{Joshua A.\ Carter}
\affil{Hubble Fellow, Harvard-Smithsonian Center for Astrophysics, 60
 Garden St., Cambridge, MA 02138}
 
  \and 
  
 \author{Eric Agol}{
 \affil{Department of Astronomy, Box 351580, University of Washington, Seattle, WA 98195, USA}





\begin{abstract}

We present a new algorithm for detecting transiting extrasolar planets in time-series photometry.  The Quasiperiodic 
Automated Transit Search (QATS) algorithm relaxes the usual assumption of strictly periodic transits by permitting a variable, but 
bounded, 
interval between successive transits.  We show that this method is capable of detecting transiting planets with significant transit timing 
variations (TTVs) without any loss of significance -- ``smearing'' -- as would be incurred with traditional algorithms; however, this is at the cost of an slightly-increased stochastic background.  
The approximate times of transit are standard products of the QATS search.  Despite the increased flexibility, 
we show that QATS has a run-time complexity that is comparable to traditional search codes and is comparably easy to 
implement.  QATS is applicable to data having a nearly uninterrupted, uniform cadence and is therefore well-suited to the 
modern class of space-based transit searches (e.g., {\it Kepler}, {\it CoRoT}).  Applications of QATS include transiting planets in 
dynamically active multi-planet systems and transiting planets in stellar binary systems.  


\end{abstract}
		
\keywords{stars: planetary systems  --- techniques: photometric --- techniques: numerical}

\section{Introduction}

This paper describes a new algorithm for detecting the transits of an extrasolar planet in a time-series of stellar flux observations.  Many algorithms have been developed and specialized for this exact purpose (e.g., \cite{2002A&A...391..369K}, \cite{2008A&A...492..617R}, \cite{2001A&A...365..330D}, \cite{2002A&A...395..625A}; see \cite{2003A&A...408L...5T} and \cite{2005A&A...437..355M} for a comparison of many of these methods). The variety of implementation in those algorithms arose primarily from the need to optimize transit detection (in the form of increased significance) in response to the nature of the data (e.g., \cite{2012MNRAS.420.1045G}, \cite{2010ApJ...713L..87J}), or to speed or automate the detection of transiting planets in a large collection of light curve data (e.g., \cite{2005ApJ...620.1033W}).  These specializations did not, however, address the potential degradation of detection significance as a result of the non-Keplerian planetary motion (exceptions are the algorithms proposed by \cite{2008MNRAS.387.1597O} and \cite{1996Icar..119..244J} to detect transiting circumbinary planets).  Specifically, all of these algorithms operate on the assumption of strict periodicity of the arrival of transits.  

Strict periodicity is not expected when greater than two bodies are interacting gravitationally (e.g., in many-planet systems or for planets in binary systems).  In these cases, we would expect non-linearities in the transit times, commonly referred to as transit timing variations (TTVs). The detection of TTVs can suggest the presence of an unseen perturbing companion (e.g., \cite{2011ApJ...743..200B}, \cite{2012Sci...336.1133N}) and yield information on the masses of the bodies involved in the gravitational exchange (\cite{2005MNRAS.359..567A}, \cite{2005Sci...307.1288H}).  The observation of TTVs in several transiting planets in a single system has proved invaluable in confirming their planetary nature and, more importantly, has placed useful constraints on their masses (\cite{2010Sci...330...51H}, \cite{2011Natur.470...53L}, \cite{2011ApJS..197....7C}, \cite{2012Sci...337..556C}).  Systems of this type are most likely to be detected by space-based transit surveys (e.g., {\em Kepler}, \cite{2010Sci...327..977B}  and {\em CoRoT}, \cite{2006cosp...36.3749B}) given their nearly continuous observing modes.

\cite{2011MNRAS.417L..16G} have shown that the failure of traditional transit algorithms to account for TTVs would result in reduction of detection significance (transit `smearing').  They showed that the  reduction in significance is non-negligible with `typical' system architectures comprised of many interacting planets in a single system.  In particular, for a transit-timing variation with an amplitude $\sigma_{TTV}$ that is longer than the transit duration and having a period shorter than the duration of the data, the signal-to-noise is reduced by a factor of $\sqrt{T_{transit}/\sigma_{TTV}}$, where $T_{transit}$ is the transit duration.  The transit-timing variation amplitude tends to grow in proportion to orbital period, $P$, while the transit duration grows as $P^{1/3}$. As such, the reduction is most significant for long period planets that are perturbed by nearby companion. 

The detection algorithm described in this paper (the Quasiperiodic Automated Transit Search or QATS) presents one possible solution to the problem raised by \cite{2011MNRAS.417L..16G} by allowing bounded variation in the intervals between transits when determining the detection significance, accounting for possible TTVs.  We pose the detection problem in \S~\ref{sec:form} and derive the algorithmic solution (QATS) in \S~\ref{sec:qats} by specializing and expanding the work by \citet{Kel'Manov522004} who tackled the more general problem of detection and pulse-shape discrimination.   In \S~\ref{sec:back}, we estimate the background significance in the presence of white noise as a function of the algorithm parameters.  In \S~\ref{sec:examples}, we provide a worked example of the application of the QATS to the {\em Kepler} data for Kepler-36 (\cite{2012Sci...337..556C}).

\section{Formulation of the problem}  \label{sec:form}

The QATS algorithm is capable of detecting the transits of an exoplanet in a series of observations of the relative flux of its host star.  

We require --- in order to simplify our discussion and to facilitate a fast algorithm --- that the $N$ observations of the normalized and median removed flux, $F(t_n)$ for $0 < n < N-1$, follow a uniform cadence; i.e., $t_n-t_0 \propto n$.  Given this, we may uniquely specify the time of any measurement by referring to its sequential 
cadence number $n$.  We use this index exclusively in what follows as opposed to referring to the absolute time and write $F(t_n)$ as $F_n$.

\subsection{Quasiperiodic transit light curve model \label{sec:setup}}

\begin{figure*}[th] 
   \centering
      \plotone{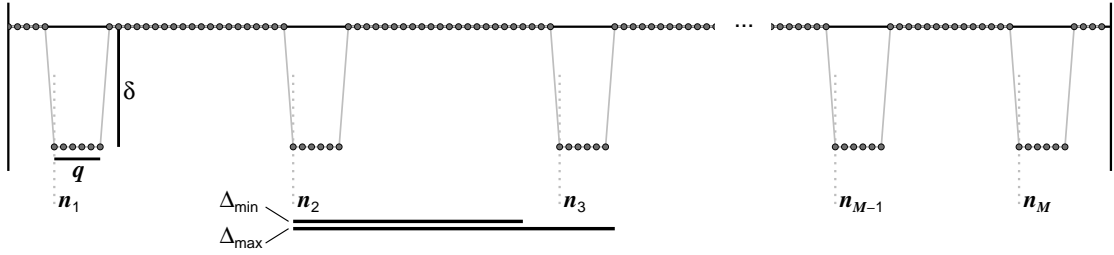}
      \caption{Cartoon representation of the QATS light curve model. Refer to the text in \S~\ref{sec:setup} for a description of the parameters shown here.}
   \label{fig:diagram}
\end{figure*}

We assume that the data are completely explained by a transit light curve model, $L(t_n) \equiv L_n$, that is contaminated by additive, independent, identically distributed Gaussian noise of width $\sigma$, such that $F_n = L_n+\epsilon$ with $\epsilon \sim {\cal N}(0,\sigma)$. Refer to Figure \ref{fig:diagram} for a qualitative representation of the model, described next.

During transit, the light curve decreases for a short amount of time when the planet occludes the star.  We make the usual
approximation (e.g., \cite{2002A&A...391..369K}) that the light curve profile of a single transit event is boxed-shaped with depth $\delta$ and duration $q$.  We assume that $q$ is an exact integral number of cadences and that the flux deficit starts precisely on a cadence.

Suppose that $M$ transits of fixed shape (fixed $q$ and $\delta$) occur in the sequence $\{L_n\}$.   Let the collection of indices at the start of each of the $M$ transits be 
\begin{equation}
\eta_M = \{n_1,...,n_{m-1},n_m,...n_{M-1},n_{M}\}.
\end{equation}   We may compactly define the model light curve sequence $\{L_n\}$ as the sum of flux deficits from all $M$ transits as
\begin{eqnarray}
	L_n & = &  \sum_{m=1}^{M} u_{n-n_m} \label{eq:model}
\end{eqnarray}
where we have defined the ``root'' transit deficit profile as
\begin{eqnarray}
	u_j & = & \left\{ \begin{array}{lr}
	  		-\delta &\;\; 0 \leq j \leq q-1 \\
			0 &\;\; \mbox{otherwise}
			\end{array}
			\right.. \label{eq:root}
\end{eqnarray} 

For a strictly periodic transiting planet, we would expect the interval, $\Delta_m$, between 
successive transits to obey (for $m > 0$)
\begin{equation}
	\Delta_m \equiv n_m-n_{m-1} = \bar{P}, \tag{Periodic Transit} \label{eq:percond}
\end{equation}
for a {\it constant} period $\bar{P}$. We have distinguished this period, measured as an integral number of cadences, from the absolute period, $P$, which is measured in units of time and may be fractional.  

We wish to relax the constancy of this interval to allow for possible transit timing variations. Fully unspecified intervals need not be considered;  we expect, based on physical considerations, that these {\it a priori} unknown variations (in interval) are {\it small} relative to the mean interval. The smallness of this variation depends on the specifics of the physical system to be detected.  For example, the amplitude of this variability is $\sim 1\%$ of the mean orbital period for known cases of interacting planets in multi-transiting systems (e.g., \cite{2010Sci...330...51H}, \cite{2011Natur.470...53L}, \cite{2012Sci...337..556C}).  For transiting circumbinary planets, the amplitude is a much larger fraction: one would expect timing variations comparable to the orbital period of the stellar binary; this may amount to a variation $\sim 10\%$ of the mean orbital period of the planet (e.g., \cite{2011Sci...333.1602D}).  In either case, we may specify a variable but nearly periodic, or `quasiperiodic' bounding, between successive transits:
\begin{equation}
	\Delta_{\rm min} \leq \Delta_m \equiv n_m-n_{m-1} \leq \Delta_{\rm max} \tag{Quasiperiodic Transit}
\end{equation}
where $\Delta_{\rm min}$ and $\Delta_{\rm max}$ are the minimum and maximum intervals considered.   Following the above, it is sometimes convenient to parameterize the difference in the maximum and minimum interval to be some small fraction, $f$, of the minimum interval: $\Delta_{\rm max}-\Delta_{\rm min} = f \Delta_{\rm min}$.

 In addition to the quasiperiodic bounding, we require that the transits be non-overlapping and that the first transit (starting at cadence $n_1$) and the last transit (starting at cadence $n_{M}$) are fully contained within the sequence.  These conditions are equivalent to
 \begin{eqnarray}
 &q& \leq \Delta_{\rm min} \label{eq:cond1} \\ 
 0 \leq &n_1&  \leq  \Delta_{\rm max}-q \label{eq:cond2} \\ 
 N-\Delta_{\rm max} \leq &n_{M}& \leq N-q. \label{eq:cond3}
\end{eqnarray}
Subject to the constraints above, the permitted number of transits in $L_n$ lies between a minimum ($M_{\rm min}$ > 1) and maximum ($M_{\rm max}$) where it is simple to show that
\begin{eqnarray}
	M_{\rm min} & = & {\rm floor}\left(\frac{N+q-1}{\Delta_{\rm max}}\right) \label{eq:mmin}\\
	M_{\rm max} & = & {\rm floor}\left(\frac{N-q}{\Delta_{\rm min}}\right) +1.  \label{eq:mmax}
\end{eqnarray}

\subsection{Maximum-likelihood objective function}

The QATS algorithm is a maximum-likelihood method.  In other words, QATS determines the set of starting cadences ($\eta_{M,{\rm best}}$), the transit duration ($q_{\rm best}$) and transit depth ($\delta_{\rm best}$) that best agrees with the data (has the maximum likelihood), subject to the quasiperiodic condition.  

We have derived an expression, more suited for numerical computation than using the full likelihood, that is also maximized for the optimal parameters. This objective function, $S(\eta_M, q)$, is independent of $\sigma$ and the transit depth when the noise is stationary.  The details of this derivation have been provided in the Appendix. The result is
\begin{eqnarray}
	S(\eta_M, q) &=&  \frac{\bar{S}(\eta_M,q) }{\sqrt{M q}}   \label{eq:final}
\end{eqnarray}
where
\begin{eqnarray}
	\bar{S}(\eta_M,q) &=& \sum_{m=1}^{M} \sum_{j=0}^{q-1}  -F_{j+n_m}.
\end{eqnarray}
The most likely transit depth (also derived in the Appendix) is
\begin{eqnarray}
	\delta_{\rm best} & = & \frac{\bar{S}(\eta_M,q) }{M q} \label{eq:deltamax}\\
		& = & \frac{S(\eta_M, q)}{\sqrt{M q}}.
\end{eqnarray}
$S(\eta_M, q)$ has a simple physical interpretation when maximized: 
\begin{eqnarray}
	\frac{S(\eta_M,q)}{\sigma} &=& \frac{\delta_{\rm best}}{\sigma} \sqrt{M q} \nonumber \\
		& \equiv &  (S/N)_{\rm total} \label{eq:SN}
\end{eqnarray}
where $(S/N)_{\rm total}$ is the total transit signal-to-noise ratio.

\section{The QATS algorithm} \label{sec:qats}

The previous section introduced an expression, $S(\eta_M, q)$ (Eqn.~\ref{eq:final}), that is maximized when the likelihood (Eqn.~\ref{eq:like}) is greatest.   The task remains to describe the algorithm that actually maximizes $S(\eta_M, q)$ with respect to the transit duration, $q$, and the set of starting cadences, $\eta_M$ for a selected $\Delta_{\rm min}$ and $\Delta_{\rm max}$ (or $\Delta_{\rm min}$ and fraction $f$). 

Plausible transit durations may be restricted to a reduced range of cadences for a given period subject to simple physical expectations (in particular, duration $\propto P^{1/3}$). In any case, the allowed range of durations may be searched, marginalizing with respect to $\eta_M$ for each duration, to find the maximum likelihood and best-fitting duration.   Similarly, the possible number of transits in an observation $\{F_n\}$ is enumerated from $M_{\rm min}$ and $M_{\rm max}$, according to Eqns.~(\ref{eq:mmin}, ~\ref{eq:mmax}), and may be searched explicitly. 



\subsection{Dynamic programming solution for maximization over $\eta_M$ by \citet{Kel'Manov522004} \label{sec:kjsol}}

The only non-trivial marginalization of $S(\eta_M, q)$ -- over $\eta_M$ for a fixed $q$ and $M$ --   is found using a specialization of the algorithm by \citet{Kel'Manov522004}.  We review this algorithm in what follows.   

First, we further distill the objective function: for a fixed $M$ and $q$, $S(\eta_M, q)$ is maximized when $\bar{S}(\eta_M,q) = \sum_{m=1}^{M} \sum_{j=0}^{q-1}  -F_{j+n_m}$ is maximized.   Note that $\bar{S}(\eta_M,q)$ may be written as
\begin{eqnarray}
	\bar{S}(\eta_M,q) & = & \sum_{m=1}^M D_{n_m} \label{eq:sbar}
\end{eqnarray}
where 
\begin{eqnarray}
	D_n = \sum_{j=0}^{q-1}  -F_{j+n}. \label{eq:dn}
\end{eqnarray}
$D_n$ can be interpreted as the discrete convolution of the data with a box of unit depth and duration $q$ (a box ``matched-filter'').  At a transit event, $D_n$ has a symmetric triangular profile of width $q$, peaking at the start of the transit in the absence of noise (see Figure \ref{fig:examp}a)).  Written this way, we see that the maximum of $\bar{S}(\eta_M,q)$ is the maximum of a sum of $M$ numbers, drawn from $D_n$ with the selected indices subject to the quasiperiodic condition.  The additive nature of this objective function along with the quasiperiodic constraints admits a `dynamic programming' solution for the maximization, to be described shortly.  Dynamic programming typically describes a class of algorithms in which calculations made at prior steps in an iteration are retained in memory for use in the current step\footnote{Classic examples of dynamic programming are those used to determine sequences defined by a recurrence relation; e.g., the Fibonacci sequence.}. 

Second, we enumerate all possible transit start times.  Viz., the first transit in the sequence $F_n$, beginning at index $n_1$, must occur before the maximum interval has passed (according to Eq. \ref{eq:cond2}).  We consider, in turn, the possible starting indices $n_1$ in this set of possible indices.  We refer to this set as $\omega_1$.  For example, with $q = 2$, $\Delta_{\rm max} =  4$, the maximum index the first transit can begin at is $n_{1, {\rm max}} = \Delta_{\rm max}-q = 2$ such that $\omega_1 = \{0,1,2\}$ and $n_1 \in \omega_1$.   The second transit, starting at index $n_2 \in \omega_2$, has to occur between $\Delta_{\rm min}$ and $\Delta_{\rm max}$ after $n_1 \in \omega_1$ and must also permit the inclusion of $M-2$ more transits, all subject to the quasi-periodic condition, in the $N$ observations.  The location of the third, fourth, etc. transit must follow analogously. Summarily, each transit $m$ of $M$ total transits is restricted to a set of possible starting indices, $\omega_m$, subject to our constraints.  \cite{Kel'Manov522004} showed that the possible sets of transit start indices may be exactly specified given $\Delta_{\rm min}$, $\Delta_{\rm max}$, $N$, $q$ and $M$ as
\begin{eqnarray}
	 \omega_m &=& \{i : n'_m \leq i \leq n''_m\} \label{eq:omegam}
\end{eqnarray}
where 
\begin{eqnarray}
	n'_m &=& {\rm max}\left[(m-1)\Delta_{\rm min}, N-\Delta_{\rm max}(M-m+1) \right] \\
	n''_m &=& {\rm min}\left[m \Delta_{\rm max}-q, N-q-(M-m)\Delta_{\rm min}\right].
\end{eqnarray}
It is also useful to consider the subset of $\omega_{m-1}$ that is conditioned on the knowledge of the starting index of the {\it next} transit.  In particular, if the $m$th transit is known to occur at index $n_m = n$, then subject to the quasiperiodic condition, the preceding transit (the $(m-1)$th transit) must have started at index $n_{m-1}$ such that $n-\Delta_{\rm max} \leq n_{m-1} \leq n-\Delta_{\rm min}$.   We refer to the intersection of the set given by this inequality and the set $\omega_{m-1}$ of allowed transit start times of the $(m-1)$th transit as $\gamma_{m-1}(n)$.  Again, \cite{Kel'Manov522004} provide the exact expression of this set in closed form:
\begin{eqnarray}
	\gamma_{m-1}(n) &  = & \{i : n'''_{m-1}(n) \leq i \leq n''''_{m-1}(n)\} \label{eq:gamma}
\end{eqnarray}
where
\begin{eqnarray}
	n'''_{m-1}(n) & = & {\rm max}[(m-2)\Delta_{\rm min}, n-\Delta_{\rm max}] \\
	n''''_{m-1}(n) & = & {\rm min}[ (m-1)\Delta_{\rm max}-q, n-\Delta_{\rm min}]
\end{eqnarray}

With the above formalism, we may detail the algorithm by \cite{Kel'Manov522004}.   We calculate the maximum of $\bar{S}(\eta_M,q)$ by considering successive transits (restricted to the indices specified by the $\omega_m$) starting with the first. At each transit number $m$ and candidate transit start index $n \in \omega_m$, we calculate an intermediate quantity $\bar{S}_{mn} \equiv \max~ \bar{S}(\eta_m | n_m = n; q)$ where $\eta_m=\left\{n_1,...,n_m\right\}$ is the subset of the first $m$ transits in $\eta_M$ which is the maximum value of $\bar{S}(\eta_m,q)$ for this circumstance, {\it i.e.\ excluding the contribution from proceeding transits but including the contribution from preceding transits}.  Given this definition and Eq.~\ref{eq:sbar}, we may recursively construct $\bar{S}_{mn}$ for all transits $m \in \{1,... M\}$:
\begin{align}
	\bar{S}_{mn} &= D_n + \max_{i \in \gamma_{m-1}(n)} \bar{S}_{{m-1},i} &\text{$n \in \omega_m$} \label{eq:smn}\\
	\bar{S}_{1n} &= D_n & \text{$n \in \omega_1$} \label{eq:s1n}
\end{align}
where we have utilized the set $\gamma_{m-1}(n)$, as defined above, in the first line.  It is then simple to show that the desired maximum of the objective function for fixed $q$ and $M$ is given by
\begin{eqnarray}
	\max_{\eta_M} \bar{S}(\eta_M,q) & = & \max_{n \in \omega_M} \bar{S}_{Mn} \label{eq:smax}
\end{eqnarray}
We calculate $\bar{S}_{mn}$ following the dynamic programming method by minimally retaining in memory  $\bar{S}_{{m-1},n}$ for all $n \in \omega_{m-1}$ so as to calculate the $m$th row of  $\bar{S}_{mn}$ without repeating any calculation.  In Fig.~\ref{fig:examp} we show a graphic interpretation of the explicit execution of this recurrence, as would be performed by a computer, for a specific example.

\begin{figure*}[th] 
   \centering
     \epsscale{1.}
      \plotone{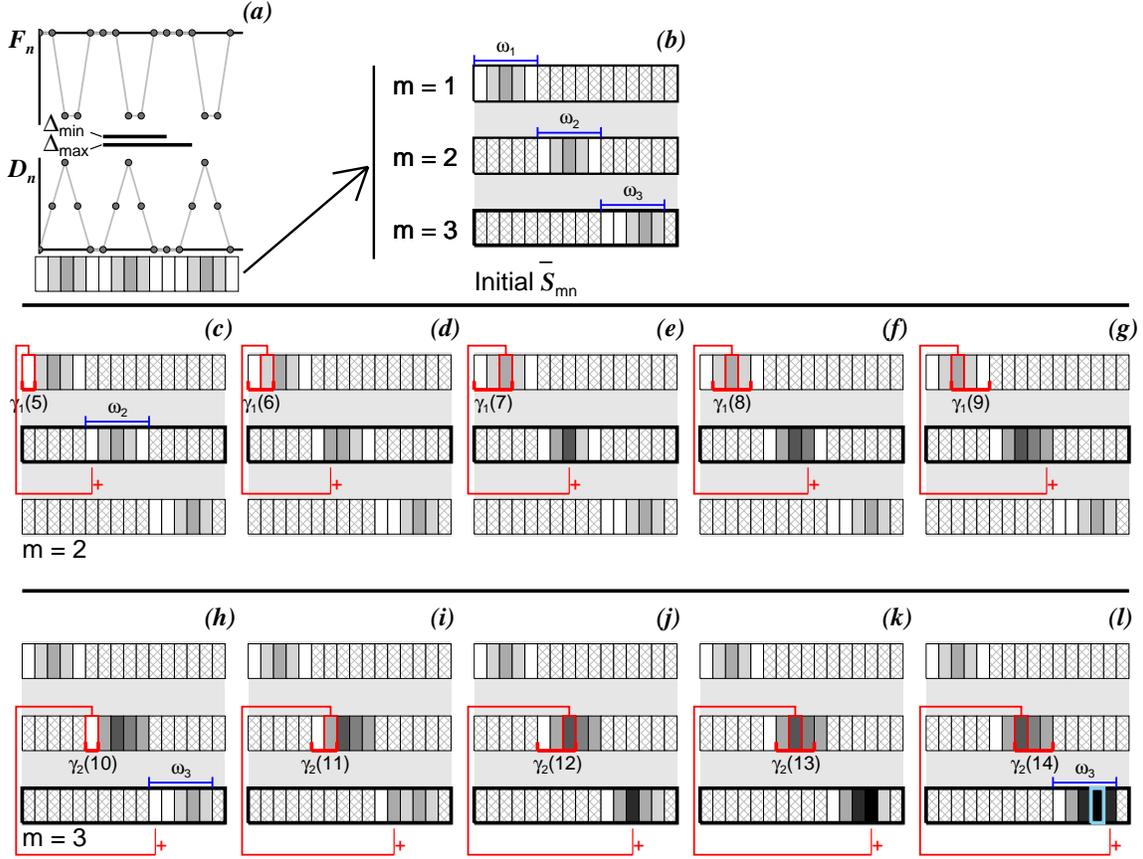}
      \caption{Here we show a graphical representation of the step-by-step execution of the QATS algorithm for a noiseless example. {\em a)} The `data' are shown in the top time-series, containing three transits of duration $q = 2$ and of quasperiodic interval.  The maximum and minimum intervals ($\Delta_{\rm min}$ and $\Delta_{\rm max}$) are shown graphically.  The box-filter has been applied to the data in the middle time-series ($D_n$, given in Eqn.~\ref{eq:dn}).  The grayscale bar represents $D_n$ as well; each cell represents a single cadence and is color coded (increasing from white to black) according to the value of $D_n$.  {\em b)} This panel shows the `initial' state of the yet-to-be-computed matrix $\bar{S}_{nm}$, described in \S~\ref{sec:kjsol}.  Each row represents an individual transit $m$ (of three total).  Each row currently has a copy of $D_n$; only those cadences without hashing in their respective cells are relevant during the computation.  The allowed set of transit start times $\omega_m$ (Eqn.~\ref{eq:omegam}) is indicated by a label and a range for each row $m$.  {\em c)} There is nothing to be done for the first row.  Here, and for {\em c)} through {\em g)} we sum the value of $D_n$ and the maximum of the significance in the first transit over the possible set of transit start times for the first transit assuming the second transit starts at the indicated location ($n = 5, 6, 7, 8$ or $9$).  These sets are shown explicitly in the first row by the braced range in red (and labelled by $\gamma_{m-1}(n)$ as defined in Eqn.~\ref{eq:gamma}). The maxima of those sets have been highlighted with a red boundary.  {\em h) -- l)} We repeat the procedure performed on the second row for the third transit iterating over the possible starts of the third transit ($n = 10,11,12,13$ or $14$). {\em l)} At this point, the $\bar{S}_{mn}$ have been calculated.  The maximum significance is then determined as the maximum of the $\bar{S}_{mn}$ with $m = 3$; that cadence has been highlighted with a blue boundary. }
   \label{fig:examp}
\end{figure*}

\subsubsection{The most likely indices at the start of the $M$ transits \label{sec:besttimes}}
The most likely set of transit start indices $\eta_M$, corresponding to $\max \bar{S}(\eta_M,q)$, may be determined once $\bar{S}_{mn}$ has been calculated by traversing this matrix in reverse\footnote{It is required to retain the entire matrix $\bar{S}_{mn}$ for this purpose.}. In detail, the index $n = n_M$ that gives the maximum value of $\bar{S}_{Mn}$ for $n \in \omega_M$ (equal to $\max_{\eta_M} \bar{S}(\eta_M,q)$ according to Eq.~\ref{eq:smax}) corresponds to the most likely starting index of the final (the $M$th) transit.  We may then recursively determine the most likely starting indices of the earlier transits conditioned on the knowledge of the location of the proceeding transit(s):
\begin{eqnarray}
	n_M & =  &\argmax_{n\in \omega_M} \bar{S}_{Mn} \label{eq:btimes1}\\ 
	n_m & = &\argmax_{n\in \gamma_m (n_{m+1})} \bar{S}_{mn} \label{eq:btimes2}
\end{eqnarray}
where $\argmax_{i \in I} a_i$ is the index $i_{\rm max}$ such that $a_{i_{\rm max}}$ is maximum for all $i\in I$ and we have made use of the conditional set $\gamma_{m-1}(n)$, defined in the preceding discussion, in the second line.  Fig.~\ref{fig:examptimes} shows this recursion for the example presented in Fig.~\ref{fig:examp}.

\begin{figure}[th] 
   \centering
     \epsscale{0.95}
      \plotone{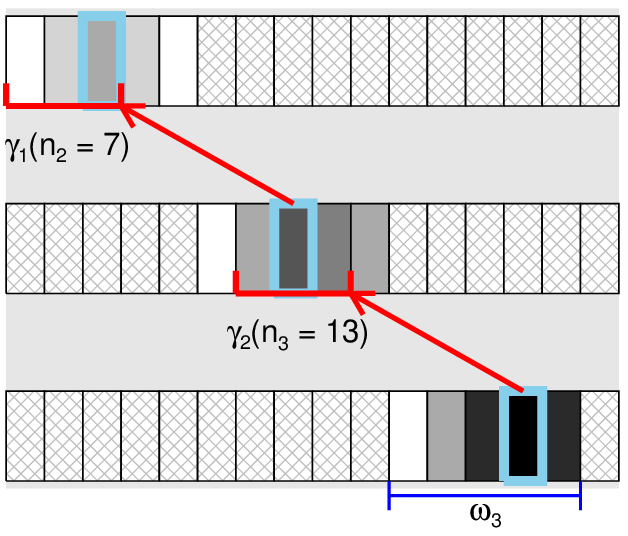}
      \caption{A graphical representation of the matrix $\bar{S}_{mn}$ and the determination of the most-likely transit start indices for the example represented in Fig.~\ref{fig:examp} according to the procedure described in \S~\ref{sec:besttimes}.  Refer to the caption of Fig.~\ref{fig:examp} for details on this graphic representation; the state of $\bar{S}_{mn}$ shown here is identical to the state in panel {\it l} of Fig.~\ref{fig:examp}. }
   \label{fig:examptimes}
\end{figure}

\subsection{QATS algorithm, in summary} 
Incorporating the solution by \cite{Kel'Manov522004}, we may summarize the QATS algorithm in pseudocode, shown as Algorithm~\ref{alg:qats}.

\begin{algorithm}[H]
\caption{QATS algorithm}
\label{alg:qats}
\begin{algorithmic}[1]
\REQUIRE $F_n$ is the data (normalized and median removed).
\STATE Given $\Delta_{\rm min}$ and $\Delta_{\rm max}$
\STATE $S_{\rm best} \gets 0$ 
\FORALL{considered transit durations $q$}
\STATE Calculate $D_n$ \COMMENT{According to Eq.~\ref{eq:dn}}
\FOR{number of transits $M = M_{\rm min}$ to $M_{\rm max}$ \COMMENT{with Eqns.~\ref{eq:mmin},~\ref{eq:mmax}}}
\STATE Calculate $\bar{S}_{mn}$ \COMMENT{According to Eqs.~\ref{eq:smn} and \ref{eq:s1n}}
\STATE $\bar{S} \gets \max_{n \in \omega_M} \bar{S}_{Mn}$
\IF{$\bar{S}/\sqrt{M q}>S_{\rm best}$}
\STATE $S_{\rm best} \gets \bar{S}/\sqrt{M q}$
\STATE $q_{\rm best} \gets q$
\STATE $M_{\rm best} \gets M$
\ENDIF
\ENDFOR
\ENDFOR
\STATE $S_{\rm best}$ is maximum of objective function with $q = q_{\rm best}$, $M = M_{\rm best}$.
\STATE Most likely transit depth is $\delta_{\rm best} = S_{\rm best}/\sqrt{M_{\rm best} q_{\rm best}}$.
\STATE Most likely transit start indices, $\eta_M = \{n_1,...,n_M\}$, calculated as in \S~\ref{sec:besttimes}.
\end{algorithmic}
\end{algorithm}
The most likely set of times may then be determined using $q_{\rm best}$ and $M_{\rm best}$ in Eqns.~\ref{eq:dn}, \ref{eq:smn}, \ref{eq:s1n} applied to Eqns.~\ref{eq:btimes1} and \ref{eq:btimes2}.

\subsection{Computational considerations \label{sec:comp}}
The QATS algorithm may be implemented as listed in Algorithm~\ref{alg:qats}, however, there are some practical considerations.  

The term $\bar{S}_{mn}$ is most simply realized in code as a $N\times M$ matrix. This matrix can be memory intensive. For example, a typical Kepler light curve, observed at the nominal $\sim30$ minute cadence and collected through observing quarter 9 (750 days), contains $N\approx$36,000 cadences.  A transiting planet with orbital period $P$ admits $M\approx750\;{\rm days}/P$ transits in that light curve. With $8$ bytes allocated per matrix entry, we can anticipate requiring $\approx 200\;{\rm MB \cdot days}/P$ of memory to store $\bar{S}_{mn}$.  Short period orbiting planets ($P < 1$ day) may require disk access, slowing the search dramatically. Additionally, when searching over multiple periods (see below), it is advantageous to conserve shared memory for the purpose of parallelization. It is therefore recommended during the calculation of $\bar{S}_{\rm best}$ that only a $2\times M$ submatrix of $\bar{S}_{mn}$ be retained  containing the $(m-1)$th row and the $m$th row under calculation.  The determination of the most likely times still requires the full matrix, however, these values are not needed for the calculation of the objective function maximum and can be calculated in a posterior operation.  

The convolved data, $D_n$, as defined in Eqn.~\ref{eq:dn}, may be modified to support `inline' data detrending with more complicated matched-filters. For example, the filter may include a linear correction outside of transit.  We also note that the $D_n$ may be pre-computed and stored in memory for a number of common durations prior to the first loop in Algorithm~\ref{alg:qats}.

It is likely that some cadences will be missing from the observation sequence (for example, those lost in a Kepler light curve during data downlink between quarters).  Ignoring these cadences violates the preconditions of the QATS algorithm and may lead to unexpected results.  We recommend filling these missing cadences with zero values.  Note in this case that transits occurring during these gaps will provide no additional significance to a detection and the most likely times of these missing transits will be meaningless.

\subsection{The QATS `spectrum' \label{sec:spec}}
The QATS algorithm, detailed in Algorithm~\ref{alg:qats}, determines the most likely duration and times for a transiting planet with `period' specified by the pair of interval bounds, $\Delta_{\rm min}$ and $\Delta_{\rm max}$.  In practice, one does not know the period of the transiting planet (nor the variation on that period between transits) and multiple periods need be compared for significance.  Traditional search algorithms establish a grid on period (typically being uniform in orbital frequency) and perform the fixed period marginalization over duration and phase at each grid point.  There is no single period in the QATS algorithm, however, a number of strategies can be suggested to perform the analogous task.  We suggest a couple below.

The first strategy is a uniform grid over minimum interval, $\Delta_{\rm min}$, with a fixed difference between this minimum interval and the maximum interval (such that $\Delta_{\rm max}-\Delta_{\rm min} = \Delta\Delta$).  In this case, a natural grid spacing is $\Delta\Delta > 0$.  In this strategy, the allowed variation in the transit interval (or period) is the same for all trial `periods' and may not reflect the underlying physical prior.  

In contrast, the second strategy allows the difference between minimum and maximum interval to be some fixed fraction, $f$, of the minimum interval (some fraction of the `period') which may better reflect our theoretical expectations \citep{2005MNRAS.359..567A} when the sought after transiting body is being perturbed by another body.  In this case, we grid $\Delta_{\rm min}$ such that the $i$th grid point is
\begin{eqnarray}
	\Delta_{\rm min}^{(i)} & = & \Delta_{\rm min}^{(0)}\left(1+f/2\right)^i \\
	\Delta_{\rm max}^{(i)} & = & \Delta_{\rm min}^{(i)}\left(1+f/2\right).
\end{eqnarray}
The fraction $f$ may be tuned for a particular search or may be searched over a small grid as well; any transit-timing variations with a fractional amplitude of $f/2$ will be contained within one of the ranges searched within this grid.

Both strategies can have coverings that are `complete' and non-overlapping over any interval in $\Delta_{\rm min}$ -- our grid resolution does not affect our ability to detect a given period for a strictly periodic transiting planet, excluding signal confusion due to a significant stochastic background signal (see \S~\ref{sec:back}).  

The detection spectrum (i.e., $\max S(\eta_M, q)$ or the maximum total transit signal-to-noise as a function of $\Delta_{\rm min}$) has characteristics similar to traditional transit search spectra that are based on a box matched-filter.  In particular, the QATS spectrum is roughly proportional to $(m n)^{-1/2}$ where $m/n$ is the rational number with smallest denominator $n$ that lies between $[\Delta_{\rm min}^{(i)}/P, \Delta_{\rm max}^{(i)}/P]$ with $P$ being the actual period of the transit. This proportionality is simply explained:  suppose your trial period is $\Delta_{\rm trial}$, while the actual
period is $P$ with $\Delta_{\rm trial}/P = m/n$ where
$m$ and $n$ are mutually prime integers. Then
every  $m$th transit will be detected, so the
signal will be reduced by a factor of $1/m$, while the noise will be changed by
a factor of $\sqrt{n/m}$: the signal-to-noise decreases by
a factor of $1/\sqrt{mn}$.  This signal-to-noise is maximized when $mn$ is
minimized, giving our dependence. The numbers $n$ and $m$ may be efficiently calculated with use of the `Farey' sequences by finding the lowest order Farey sequence which has a fraction, $m/n$, contained within the trial period search range (in units of the correct orbital period);  this guarantees that $m/n$ is the smallest mutually prime ratio and thus has the highest signal-to-noise relative to the peak.  For trial periods above the correct period, the Farey sequence search should take place between $[P/\Delta_{\rm min}^{(i)}, P/\Delta_{\rm max}^{(i)}]$.  Figure \ref{fig:farey} shows a plot of $1/\sqrt{mn}$ without the presence of noise for an interval width of $f=0.001$.  In practice, we find that this pattern matches the QATS spectrum of detected planet candidates well, while other forms of light curve variability show a different shaped QATS spectra. We note that the robustness of a detection of a transiting planet (or at least a periodic box signal) may be validated by comparing the measured spectrum to this prediction for a specified candidate period.  See \S~\ref{sec:examples} for example QATS spectra computed from noisy data.

\begin{figure}[th] 
   \centering
     \epsscale{1.1}
      \plotone{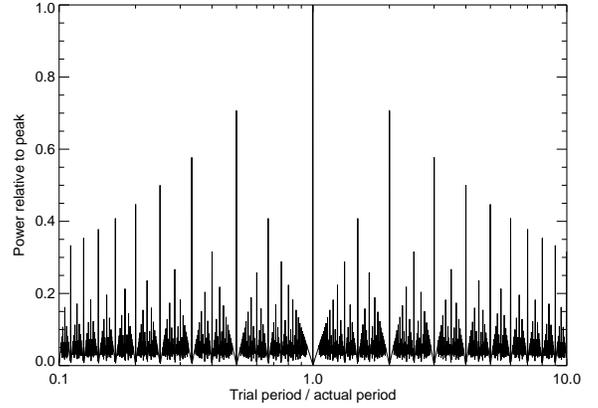}
      \caption{Noise-free expectation for the shape of the QATS spectrum for $f=\Delta_{\rm max}/\Delta_{\rm min}-1 =$0.1\%.}
         \label{fig:farey}
\end{figure}

\subsubsection{Stochastic background \label{sec:back}}

The QATS algorithm permits the detection of transiting planets with large variations in transit interval without suppressing the significance of the transit; however, the exchangeable cost of this freedom is increased backgrounds that may `swamp' a potential detection signal.  The stochastic background  -- i.e., the conspiracy of noise that yields a smooth level of non-zero significance beneath the sharp transit detections -- grows as the difference between $\Delta_{\rm max}$ and $\Delta_{\rm min}$ widens.  It is difficult to obtain an exact closed form expression for this background; however, in what follows we derive an upper bound with some simple assumptions and, further, propose formulae that closely approximate the true behavior of the background.

We assume a white-noise background, $F_n \sim {\cal N}(0, \sigma^2)$; i.e., the data are independent, identically distributed Gaussian variables with zero mean and variance $\sigma^2$.  We consider the distribution of the maximum of the total transit signal-to-noise of this background where, according to Eqns.~\ref{eq:SN}, \ref{eq:final} and \ref{eq:sbar},
\begin{eqnarray}
	\max (S/N)_{\rm total,BG} & = & \max_{\eta_M} \frac{S(\eta_M,q)}{\sigma} \nonumber \\
	& = & \max_{\eta_M} \frac{\bar{S}(\eta_M,q)}{ \sqrt{M q \sigma^2}} \nonumber  \\
	 &=&  \max_{\eta_M} \sum_{m=1}^M D_{n_m} \left(M q \sigma^2\right)^{-1/2} \\
	 & = & \max_{\eta_M}\sum_{m=1}^M D'_{n_m} M^{-1/2} \label{eq:snbg} 
\end{eqnarray}
Referencing Eqn.~\ref{eq:dn}, we see that the $D_n \left(q \sigma^2\right)^{-1/2} \equiv D'_n$ are Gaussian distributed with zero mean and unit variance and are correlated whenever $|n_{m_1}-n_{m_2}| < q$.  The requirement that the transits be non-overlapping (Eqn.~\ref{eq:cond1}) implies that the $\{D'_{n_m}\}$ are independent.  As such, it is simple to see that the sum $\Sigma_{\eta_M} \equiv \sum_{m=1}^M D'_{n_m} M^{-1/2}$ is distributed identically as the $D'_n$; these variables are also correlated as discussed below.

We may then regard the maximization of $\Sigma_{\eta_M}$ over $\eta_M$ as being approximately equal to the maximum, $X$, of some appropriate number of {\em independent} draws $K$ from a normal distribution with zero mean and unit variance.  The number $K$ represents the effective number of independent transit configurations, yielding independent sums $\Sigma_{\eta_M}$.  It can be shown (\cite{cramer}) that the mean of the extremum variable $X$ is asymptotically approximated for large $K$ as
\begin{eqnarray}
	\langle X \rangle_{K} & = & \sqrt{2 \log K}- \frac{\log \log K + \log 4 \pi}{2\sqrt{2 \log K}}+O\left(\frac{1}{\sqrt{\log K}}\right)\;\;\;\mbox{for $K \gg 1$}. \label{eq:meanL}
\end{eqnarray}

When $\Delta_{\rm max} = \Delta_{\rm min}$ (i.e., periodic transits), the number of draws $K$ is equal to the number of independent choices for the start time of the first transit as there is no additional freedom in the choices of the remaining $M-1$ transits.  There are  $K = \Delta_{\rm max}-q+1$ available cadences for the start of the first transit; however, for $q \neq 1$, these choices are not independent (with correlation length $\sim q$).  We can account for this {\em over-counting} by dividing $K$ by an effective correlation length $l = q/\alpha$ for some $\alpha \geq 1$.  Then,  the number of independent sums is $K_{\rm periodic} \approx \alpha \Delta_{\rm max}/q$.  Referring to Eqn.~\ref{eq:meanL}, for $\Delta_{\rm max}$ large, we have that
\begin{equation}
	\max (S/N)_{\rm total,BG} \approx \sqrt{2 \log  \left(\alpha \Delta_{\rm max}/q\right)} \tag{Periodic Search}.
\end{equation}
This analytic theory compares well with simulation (see Fig.~\ref{fig:bg}) so long as $\alpha = \alpha(q)$; $\alpha = \{1.0, 2.2, 3.0\}$ gives excellent agreement with simulated periodic QATS backgrounds when $q = \{1, 5, 14\}$, respectively. Obviously, this dependence also applies to singleton transits for any choice of $\Delta_{\rm max} > \Delta_{\rm min}$. 

 We may similarly consider the total number of ways to admit $M$ transits in $F_n$ for $\Delta_{\rm max} > \Delta_{\rm min}$.  As before, the first transit may start at $\Delta_{\rm max}-q+1 $ positions.  Following the choice of the $m$th transit, the following $(m+1)$th transit must start at a choice of $\Delta_{\rm max}-\Delta_{\rm min}+1$ positions, according to the quasiperiodic condition.  Applying the rule of products, and accounting for the correlation length at each transit, the total possible transit configurations is therefore 
 \begin{eqnarray}
 	K_{\rm Total} &\approx& \left(\frac{\alpha}{q}\right) \Delta_{\rm max}\prod_{m=2}^{M} \left(\frac{\alpha}{q}\right) \left(\Delta_{\rm max}-\Delta_{\rm min}+1\right)  \nonumber  \\
		& = & \left(\frac{\alpha}{q}\right)^M \Delta_{\rm max} \left(\Delta_{\rm max}-\Delta_{\rm min}+1\right)^{M-1} 
\end{eqnarray}
However, it must be the case that $\max (S/N)_{\rm total,BG} \leq \langle X \rangle_{K_{\rm Total}}$ as the $K_{\rm total}$ choices are not {\it independent} choices in this problem.  This is because the contribution to the {\it maximized} sum, $\Sigma_{\eta_M}$, is greater for $m$ small than that for $m$ large.   We may attempt to account for this variable contribution by weighting the above product by a simple function of transit number $m$:
 \begin{eqnarray}
 	K_{\rm QATS} &\approx&  \left(\frac{\alpha}{q}\right) \Delta_{\rm max}\prod_{m=2}^{M} \left(\frac{\alpha}{q}\right) \left(\Delta_{\rm max}-\Delta_{\rm min}+1\right) m^{-1/r}  \nonumber  \\
		& = &\left(\frac{\alpha}{q}\right)^M \Delta_{\rm max} \left(\Delta_{\rm max}-\Delta_{\rm min}+1\right)^{M-1}  \left[(M-1)!\right]^{-1/r}
\end{eqnarray}
where $1/r$ is a constant shaping parameter.  We then expect the background signal in the QATS search to have a dependence according to $\max (S/N)_{\rm total,BG} = \langle X \rangle_{K}$ (Eqn.~\ref{eq:meanL}) with $K = K_{\rm QATS}$ where, to simplify the computation, in the typical limit of $K_{\rm QATS} \gg 1$,
\begin{eqnarray}
	\log K_{\rm QATS} &\approx& -M\log \frac{q}{\alpha} + \log \Delta_{\rm max}+(M-1)\log\left(\Delta_{\rm max}-\Delta_{\rm min}+1\right)\nonumber \\
		&& -\frac{1}{r} \left[ (M-1) \log (M-1) - (M-1)\right]. \label{eq:qatsL}
\end{eqnarray}
In Fig.~\ref{fig:bg}, we plot the analytic prediction for the background as a function of $\Delta_{\rm max}/q$ for a selection of $\Delta_{\rm max}-\Delta_{\rm min}$, in units of duration $q$, along with simulated backgrounds of the same parameters using the QATS algorithm.  In this figure, we have assumed $M \approx (N/q)/(\Delta_{\rm max}/q)$ with $N/q = 2500$ which is reasonable in a QATS search for transits of 7 hour duration in a long cadence Kepler light curve spanning 750 days ($N = 35000$, $q = 14$). 

We find that $r$ is dependent on $\Delta_{\rm max}-\Delta_{\rm min}$ with $r \approx 11.76$, $7.14$ and $6.25$ giving the closest agreement between simulation and theory for $\Delta_{\rm max}-\Delta_{\rm min} = 1q$, $2 q$ and $3 q$, respectively.  These choices of $r$ (and those for $\alpha = \alpha(q)$ as provided above) have been used in generating the theory curves in Fig.~\ref{fig:bg}.  

The background level may be compared to the detection threshold of 7.1 set by the Kepler mission (\cite{2010ApJ...713L..87J}), for example.  The background exceeds this threshold for short periods, $\Delta_{\rm max}/q \lesssim 40$, when  $\Delta_{\rm max}-\Delta_{\rm min} = 1q$. For the Kepler-specific example, this corresponds to absolute periods of $\lesssim 10$ days with quasiperiodic variability $f \gtrsim 2.5\%$ in $750$ days of long cadence data.

The above formulae are useful to gauge the growth of the background QATS signal in presence of white noise, however, the actual background in a specific search is best represented via simulation if possible.  

The starting indices of the most likely transit signal in the presence of background alone are best described by a one dimensional constrained random walk whose intervals are approximately uniform between the minimum and maximum intervals.  This characteristic is specific to random noise and may be diagnostic of such; we would generically expect physical timing variations to have a non-stochastic form.

\begin{figure*}[th] 
   \centering
     \epsscale{1.}
      \plotone{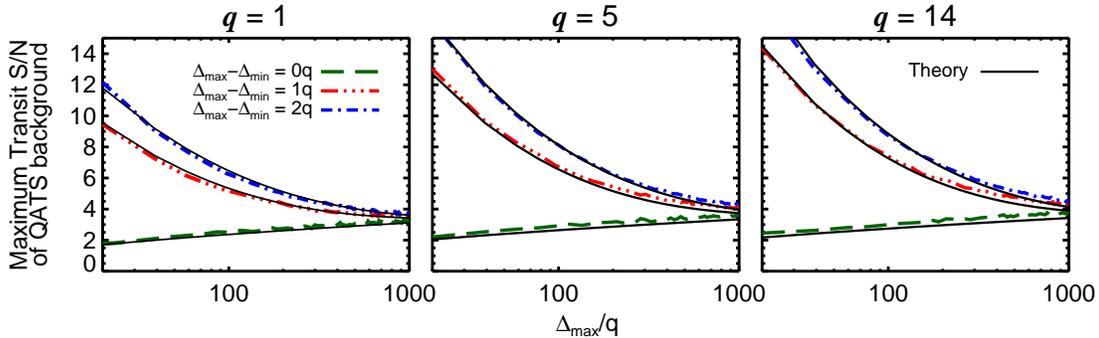}
      \caption{Stochastic background of the QATS search as a function of maximum search interval (normalized by transit duration) in the presence of `white' noise.  The broken lines show the background total transit signal-to-noise significance as determined from a Monte Carlo simulation using the QATS algorithm for a selection of transit durations $q$ and $\Delta_{\rm max}-\Delta_{\rm min}$.  The solid lines plot theory curves, given by Eqn.~\ref{eq:meanL} with $K$ given as in Eqn.~\ref{eq:qatsL}, for the parameters described in \S~\ref{sec:back}. }
         \label{fig:bg}
\end{figure*}

\subsection{Run-time analysis \label{sec:runtime}}

The success of the QATS algorithm is contingent on being both precise and efficient in computational time (and secondarily in memory-usage, see \S~\ref{sec:comp}).  In the case of the latter, the QATS algorithm performs its most complex computational task and the only task that is distinct from fixed-period searches, that of computing $\bar{S}_{mn}$ (described in \S~\ref{sec:kjsol}), in `polynomial' time-complexity.  In other words, the number of simple computational evaluations on a computer scales as some (small) power of the parameters of the problem.  In particular,  for fixed $q$ and $M$, it can be shown (\cite{Kel'Manov522004}) that the computation of $\bar{S}_{mn}$ and the maximization of the final row (see \S~\ref{sec:kjsol}) has time-complexity
\begin{eqnarray}
	C = \left\{\begin{array}{lr}
      O\left[ M \left(\Delta_{\rm max}-\Delta_{\rm min}+q\right)\left(N-q+1\right) \right]& M \geq 2 \\
      O\left[ q(N-q+1) \right]&M = 1
     \end{array}
     \right.
\end{eqnarray}
The time to execute a search scales only linearly with the (QATS-specific) interval difference $\Delta_{\rm max}-\Delta_{\rm min}$.   For the most typical case of $\Delta_{\rm max}-\Delta_{\rm min}$ small (see \S~\ref{sec:setup}),  the computational penalty of applying the QATS algorithm as opposed to a traditional fixed-period search is negligible.

\section{Example application: Kepler-36 } \label{sec:examples}

In this section, we present a complete application of the QATS algorithm on Kepler light curve data for the confirmed planetary system Kepler-36 (\cite{2012Sci...337..556C}, planet b also detected by \cite{2012arXiv1206.5347O} as KOI-277.02).  

Kepler-36 is a planetary system with two confirmed transiting planets with orbital periods of approximately 13.84 days and 16.23 days.  The proximity of these planets' orbits permits significant gravitational interactions at their closest approach which modify their orbits sufficiently to yield significant (many hour) piecewise non-linearities in both their times of transit.  As may be expected (see, e.g., \cite{2012ApJ...750..113F}), the timing non-linearities are anti-correlated between the two planets. 

The planets' radii differ by a factor of nearly 2.5 such that the transit of the smaller, interior planet b results in a loss of stellar flux that is only $\approx17\%$ that caused by the transit of the more massive planet c;  while the transits of planet c are significant by eye, those due to b are not.  While both planets may be detected with a fixed-period search (\cite{2012arXiv1206.5347O}), the QATS algorithm produces detections at much higher significance (by factors $\sim 1.7$) and provides transit times that can be seen immediately to be anti-correlated without the need for any further analysis.

\subsection{Preparation of the Kepler data for Kepler-36}

The available Kepler data for Kepler-36 span approximately 877 days of nearly uninterrupted observation at the 29.4 minute long cadence interval (equivalent to 43,053 cadences).  We utilize the `raw' aperture photometry light curve (SAP\_AP\_FLUX in the Kepler {\it fits} product, see \cite{2010ApJ...713L..87J}). A simple moving average of 50 cadences in width ($\approx 1$ day) was divided through the data to remove astrophysical and systematic trends.  The detrended light curve was then median subtracted and missing cadences were filled with zeroes.  The resulting light curve is shown in the top panel of Fig.~\ref{fig:ex:lc}; the lower panel shows the light curve after removing the readily visible transits of planet c and replacing those transit cadences with zeroes.  We assume the data have point-to-point scatter $\sigma = 7.89\times10^{-5}$ according to \cite{2012Sci...337..556C}.

\begin{figure*}[th] 
   \centering
     \epsscale{1.0}
      \plotone{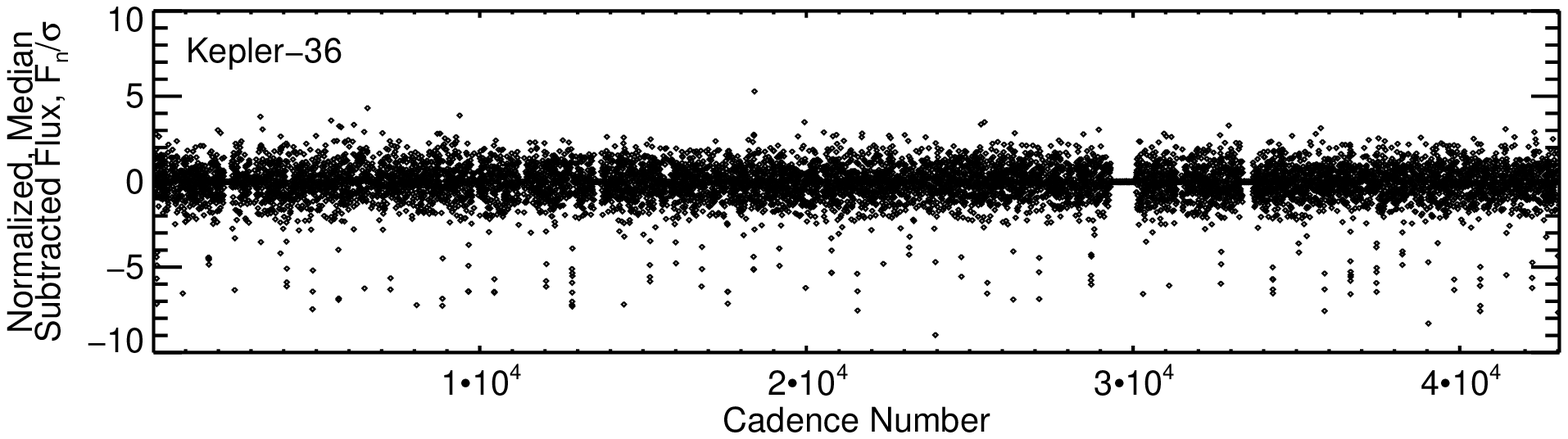}
      \plotone{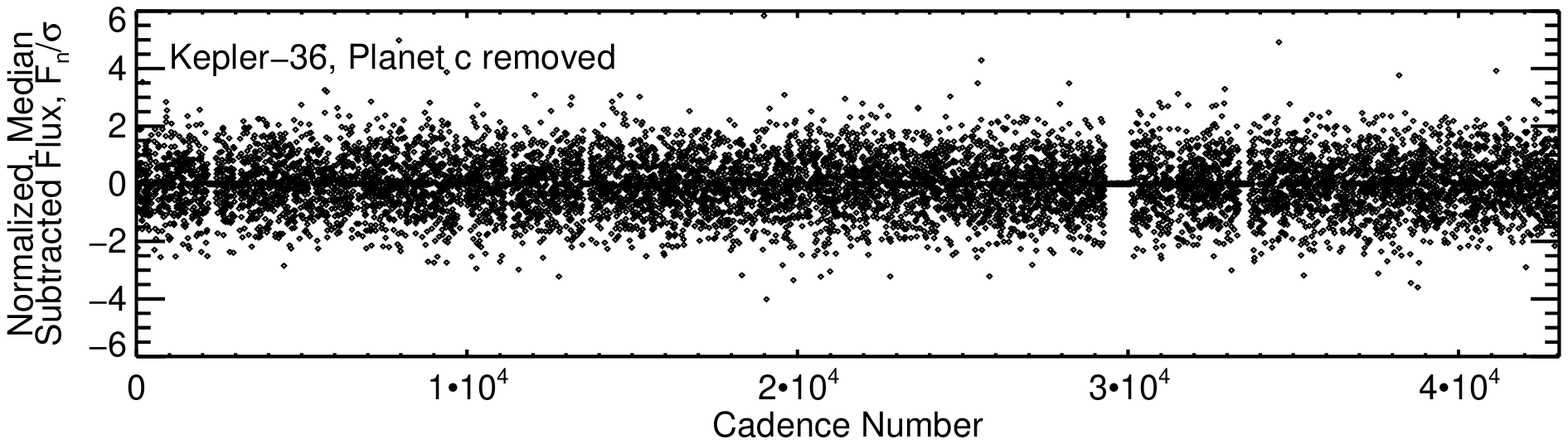}
      \caption{{\em Kepler} data for Kepler-36.  The top panel shows the available {\em Kepler} light curve data (plotted as a function of cadence number as opposed to time) for Kepler-36 after detrending and subtracting the median level (see \S~\ref{sec:examples} for details).  Missing cadences have been filled with zero values.  The transits of planet c are evident; those of planet b are not.  The bottom panel shows the same data after replacing the cadences when planet c is in transit with zeroes. }
         \label{fig:ex:lc}
\end{figure*}

\subsection{The detection of planet c}

We first apply the QATS algorithm to the light curve shown in the top of Fig.~\ref{fig:ex:lc}, containing the transits of both planets b and c.  We execute the QATS algorithm outlined in Algorithm \ref{alg:qats} for each $\Delta_{{\rm min},i}$ in a grid where $\Delta_{{\rm min},i} = i$ for $0 \leq i \leq$ $43,052$ and for three different quasiperiodic constraints $\Delta_{{\rm max},i} - \Delta_{{\rm min},i} = \{0, 1, 2\}$.  In other words, we find the maximum total transit signal-to-noise, $\max S(\eta_M,q)/\sigma$, for all minimum intervals (in integer cadence) plausible for detection in the available data.   We also fix the transit duration to $q = 14$, close to the duration, measured in cadences, as reported by \cite{2012Sci...337..556C} for planet c.

We plot in the top of Fig.~\ref{fig:ex:crv} $\max S(\eta_M,q)/\sigma$ as a function of $\Delta_{\rm min}$ for each choice of quasiperiodic constraint.  The profile of the QATS spectrum is dominated by the transits of planet c and is similar for all three choices (with profile explained analytically as described in \S~\ref{sec:spec}). However, the significance of the detection peak (around 794 cadences or $\approx$ 16.2 days) and the level of the stochastic background (see \S~\ref{sec:back}) increase with increasing $\Delta_{{\rm max},i} - \Delta_{{\rm min},i}$.  In particular, we find that the detection significance of the transits of planet c increases from $\approx 80$ to $\approx 130$ with the most dramatic improvement going from strictly periodic transits ($\Delta_{{\rm max},i} - \Delta_{{\rm min},i} = 0$) to that with the freedom of single cadence between subsequent transits. This improvement in significance coincides with better estimates of the instants of the transits (as shown with the progression of `riverplot' figures in the bottom panel of Fig.~\ref{fig:ex:crv}) resulting in less dilution or smearing of the transits by the featureless light curve baseline. 

\begin{figure*}[th] 
   \centering
   \epsscale{1.0}
   \plotone{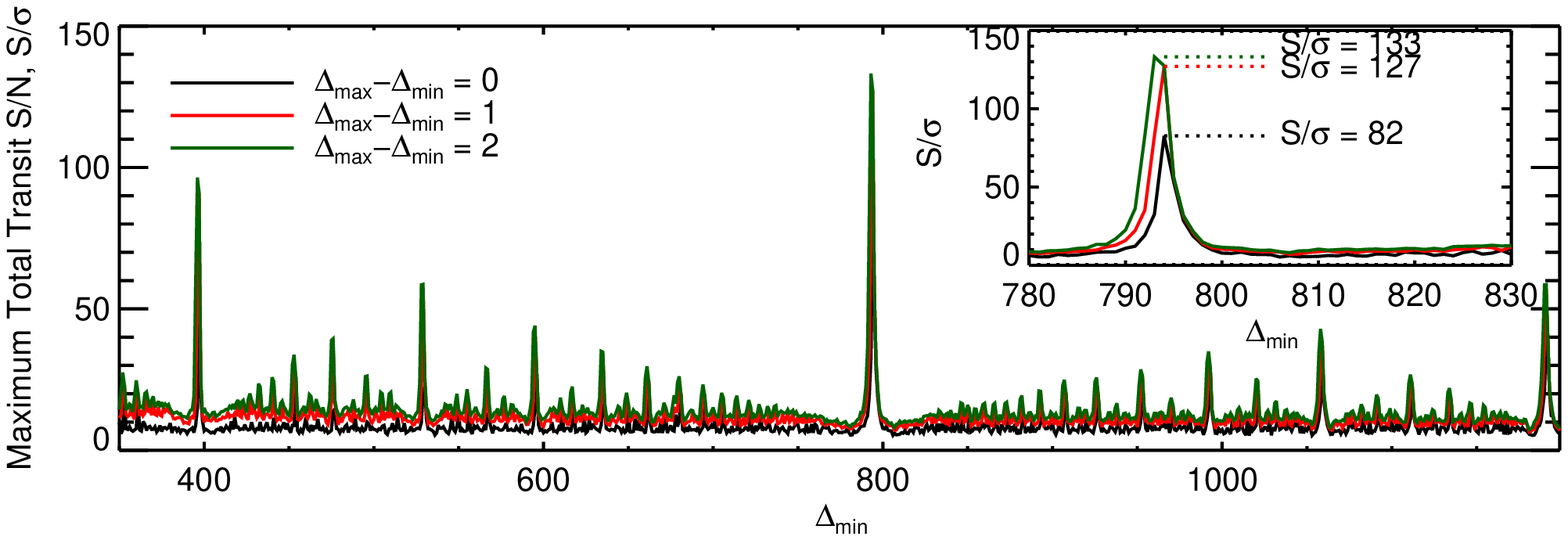}
     \epsscale{0.35}
      \plotone{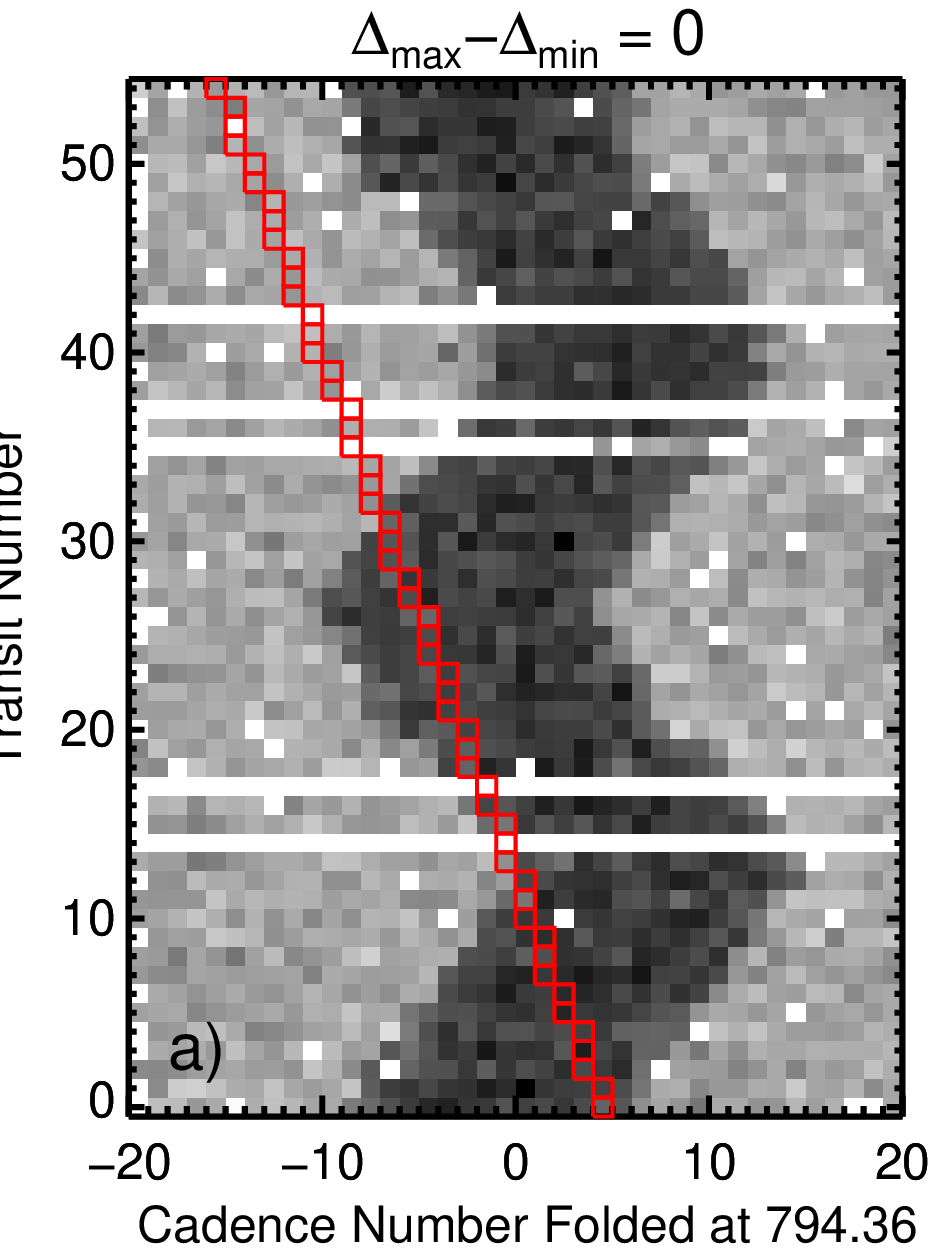}
      \plotone{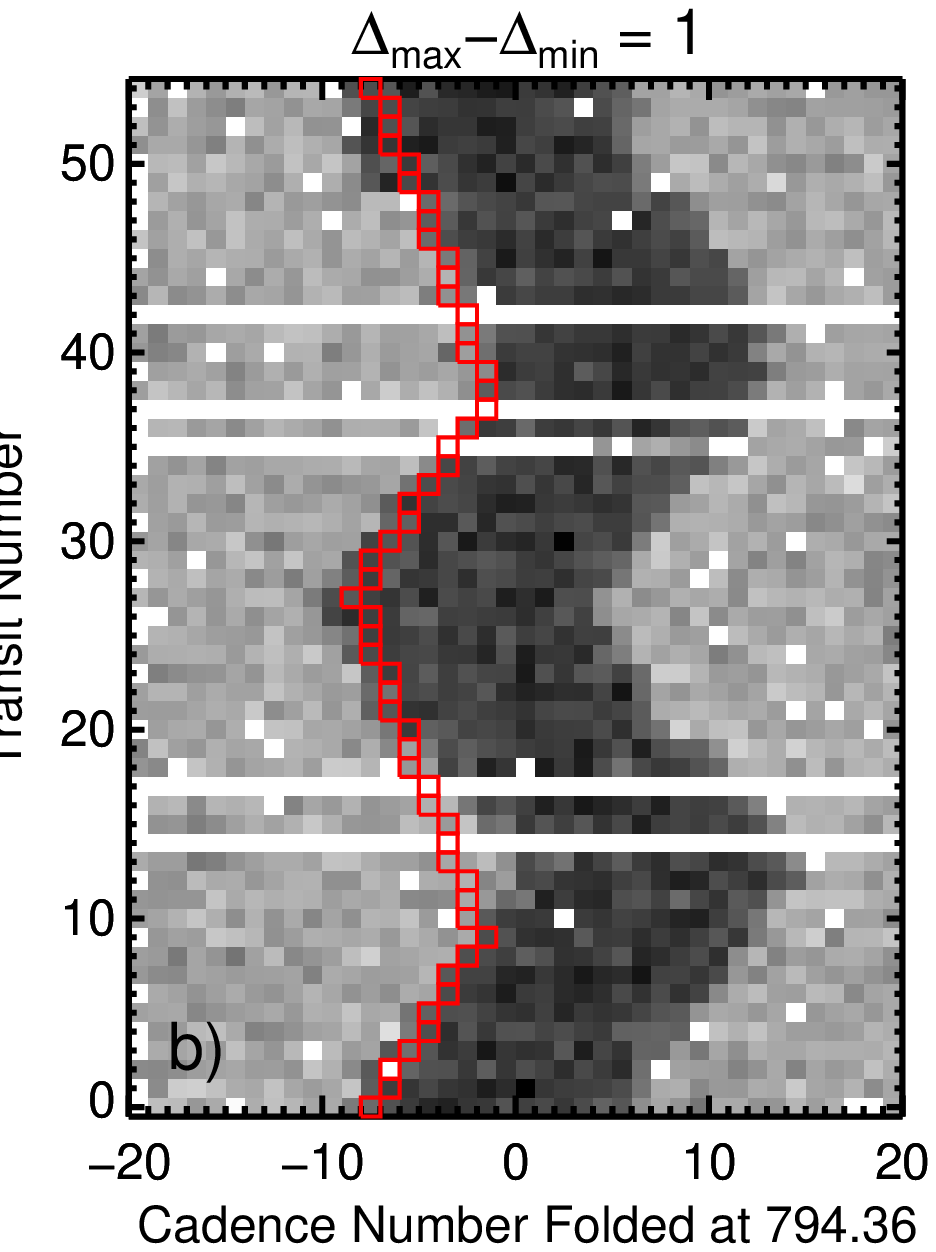}
      \plotone{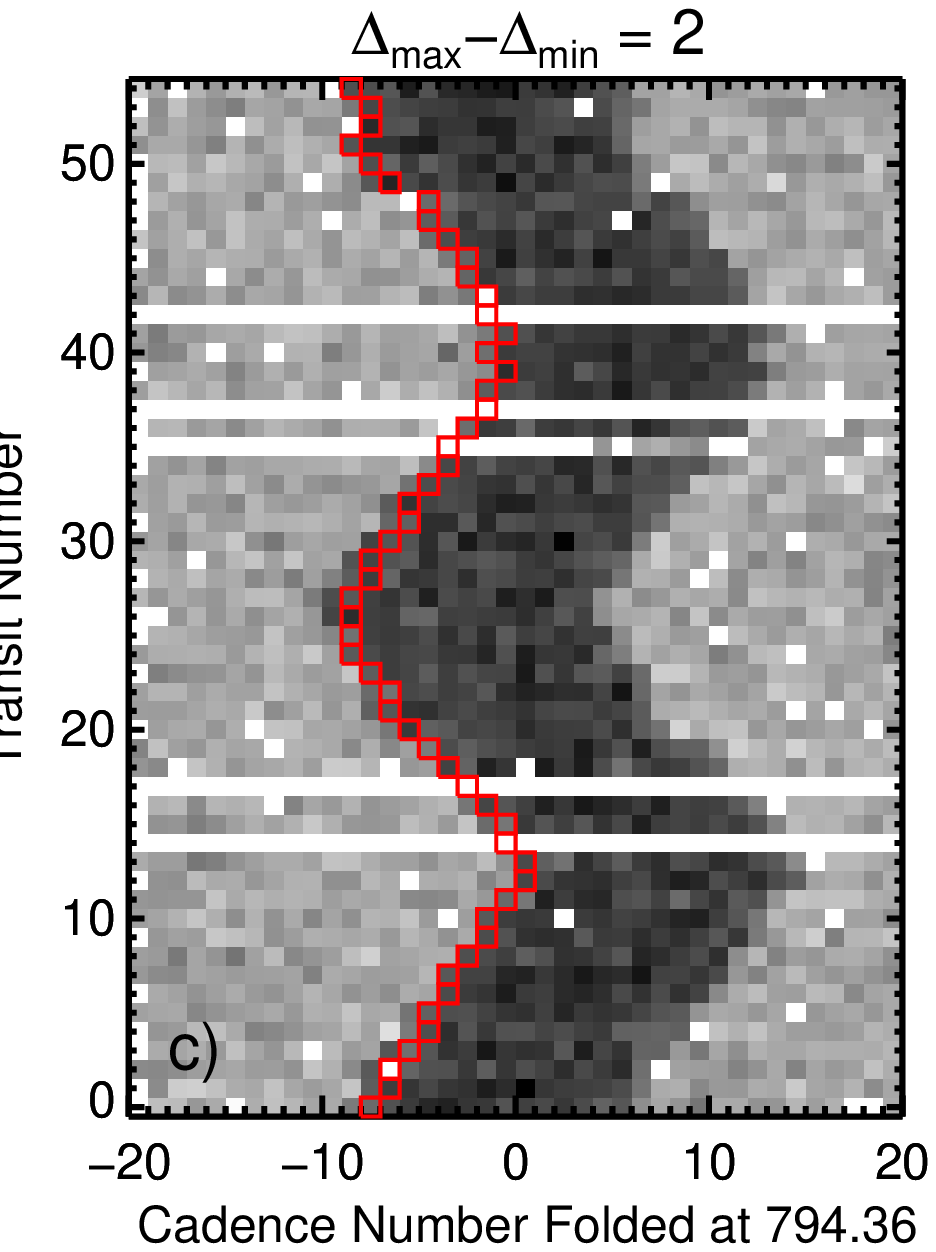}
      \caption{The detection of Kepler-36c.  {\em Top}. The top panel shows the QATS `spectrum,' maximum detection significance as a function of minimum interval (in cadences) $\Delta_{\rm min}$, for the data shown in the top panel of Fig.~\ref{fig:ex:lc} for a selection of fixed $\Delta_{\rm max}-\Delta_{\rm min}$.  The highest significance peak corresponds to the period of planet c (again in cadences; each cadence is 29.4 minutes long, the period of planet c is $\approx 794$ cadences or $\approx 16.2$ days).  The profile of the spectrum closely resembles that described analytically in \S~\ref{sec:spec}.  The significance of the detection peak increases with increasing $\Delta_{\rm max}-\Delta_{\rm min}$ (as does the background).  {\em Bottom panels}.  These 'river plots,' normalized stellar flux as a function of transit number and time modulo the mean orbital period of planet c, correspond to the QATS spectra in the top panel.  The cadences highlighted in red give the most likely instants of the transit at each transit number according the QATS algorithm.  More freedom in the quasiperiodic constraint permits more accurate determination of the transit start times (and a subsequent boost to the detection significance). 
       }
         \label{fig:ex:crv}
\end{figure*}

\subsection{The detection of planet b}

After removing the transits of planet c (having located their starts in the previous QATS application with $\Delta_{{\rm max},i} - \Delta_{{\rm min},i} = 2$), we perform a subsequent QATS search under the same conditions so as to identify the transits of planet b.  We plot in the top of Fig.~\ref{fig:ex:brv} the maximum signal-to-noise as a function of a (smaller range) of $\Delta_{\rm min}$ for $\Delta_{{\rm max},i} - \Delta_{{\rm min},i} = \{0, 1, 2\}$.  The peak in this spectrum corresponding to planet b (at $\Delta_{\rm min} = 677$) is the only dominant feature in this secondary spectrum outside of the sloped stochastic background.  The background, in this case, is appreciable compared to the detection peak, again growing with increasing quasiperiodic interval.  The detection peak remains clear (despite this background) growing from $\approx 10$ for periodic transits to $\approx 17$ for the most liberal constraint of $\Delta_{{\rm max},i} - \Delta_{{\rm min},i} = 2$.   The significance of individual transits of planet b are close to 3, making their visual identification difficult -- even when folded at the best-fitting linear ephemeris (see the bottom panels of Fig.~\ref{fig:ex:brv}).  However, with the guidance of the most-likely transit instants in the final figure of the lower panel of Fig.~\ref{fig:ex:brv}, its presence becomes clear.  This is made even more compelling given that the transit times are anti-correlated with those of planet c (compare the final figures in Figs.~\ref{fig:ex:crv} and \ref{fig:ex:brv}).

A subsequent execution of the QATS algorithm, having now removed the transits of planet b, rendered no additional detections exceeding a threshold of 7.1.

\begin{figure*}[th] 
   \centering
   \epsscale{1.0}
   \plotone{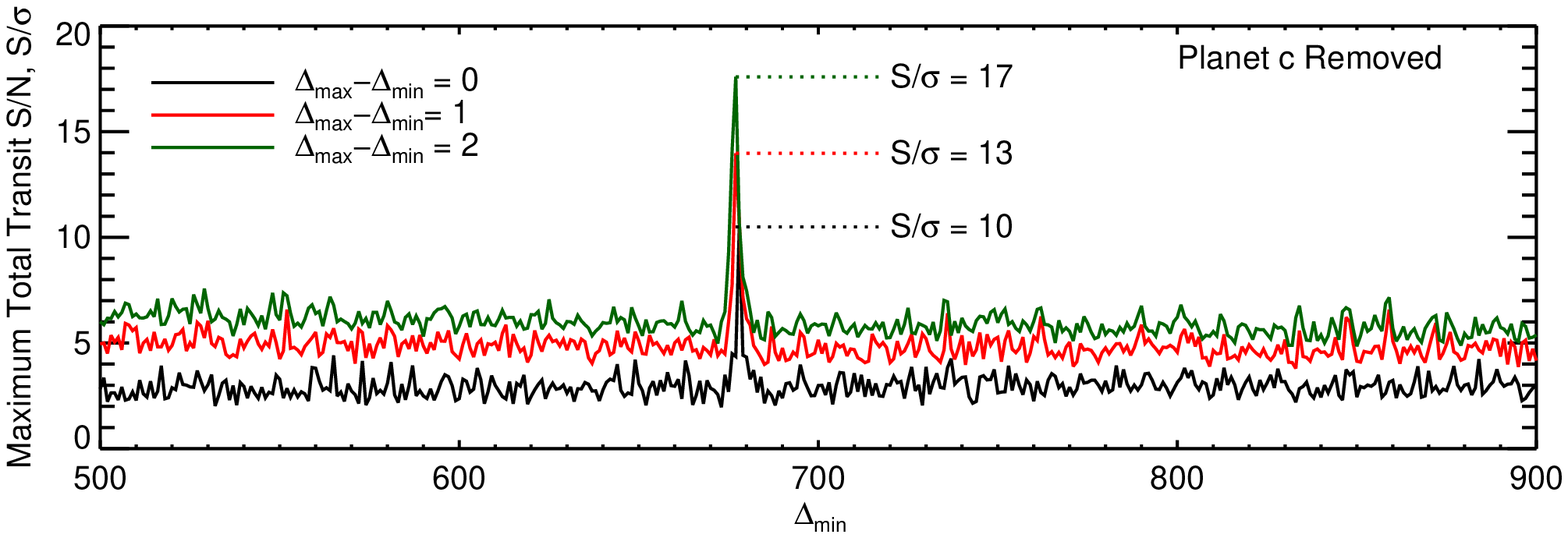}
     \epsscale{0.35}
      \plotone{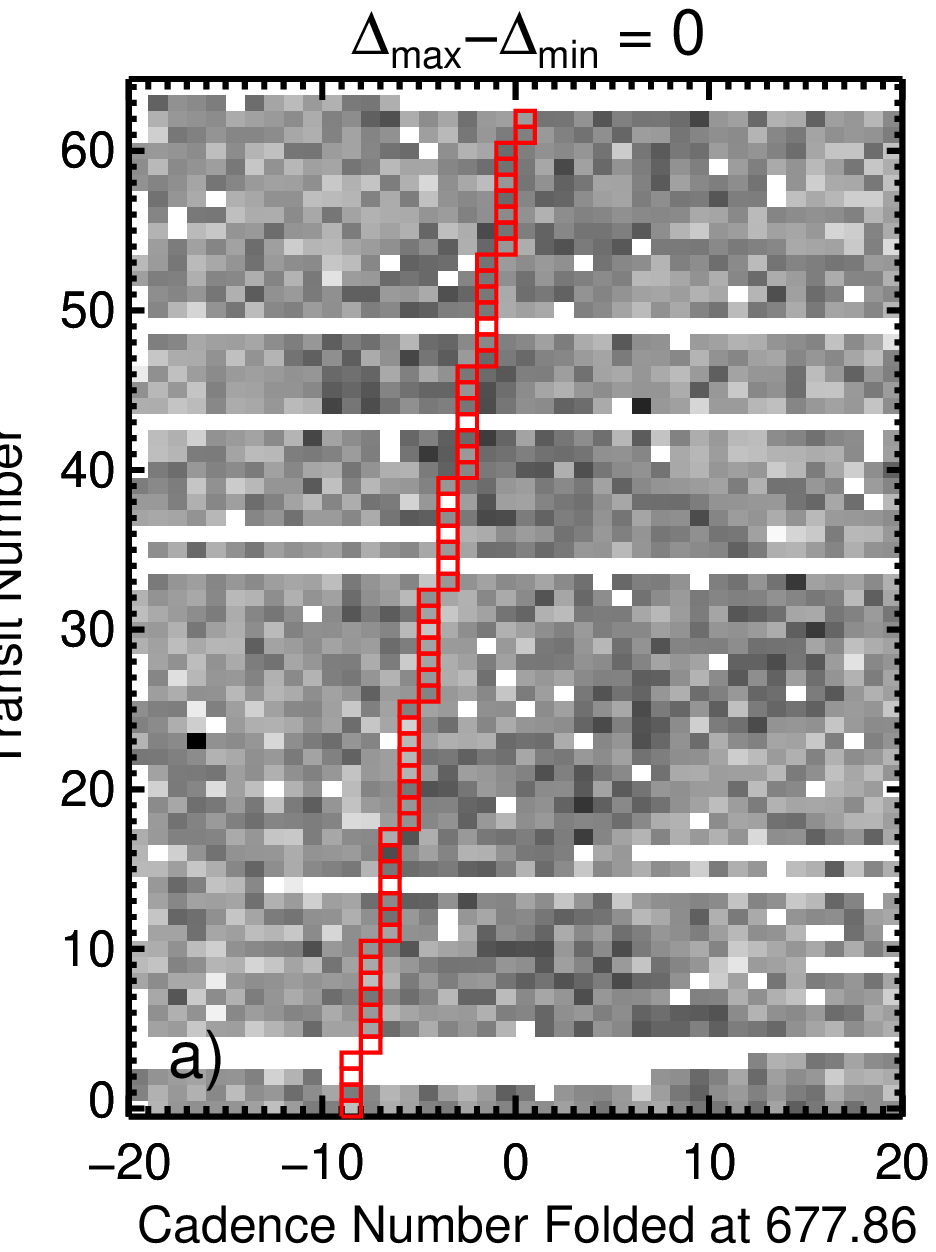}
      \plotone{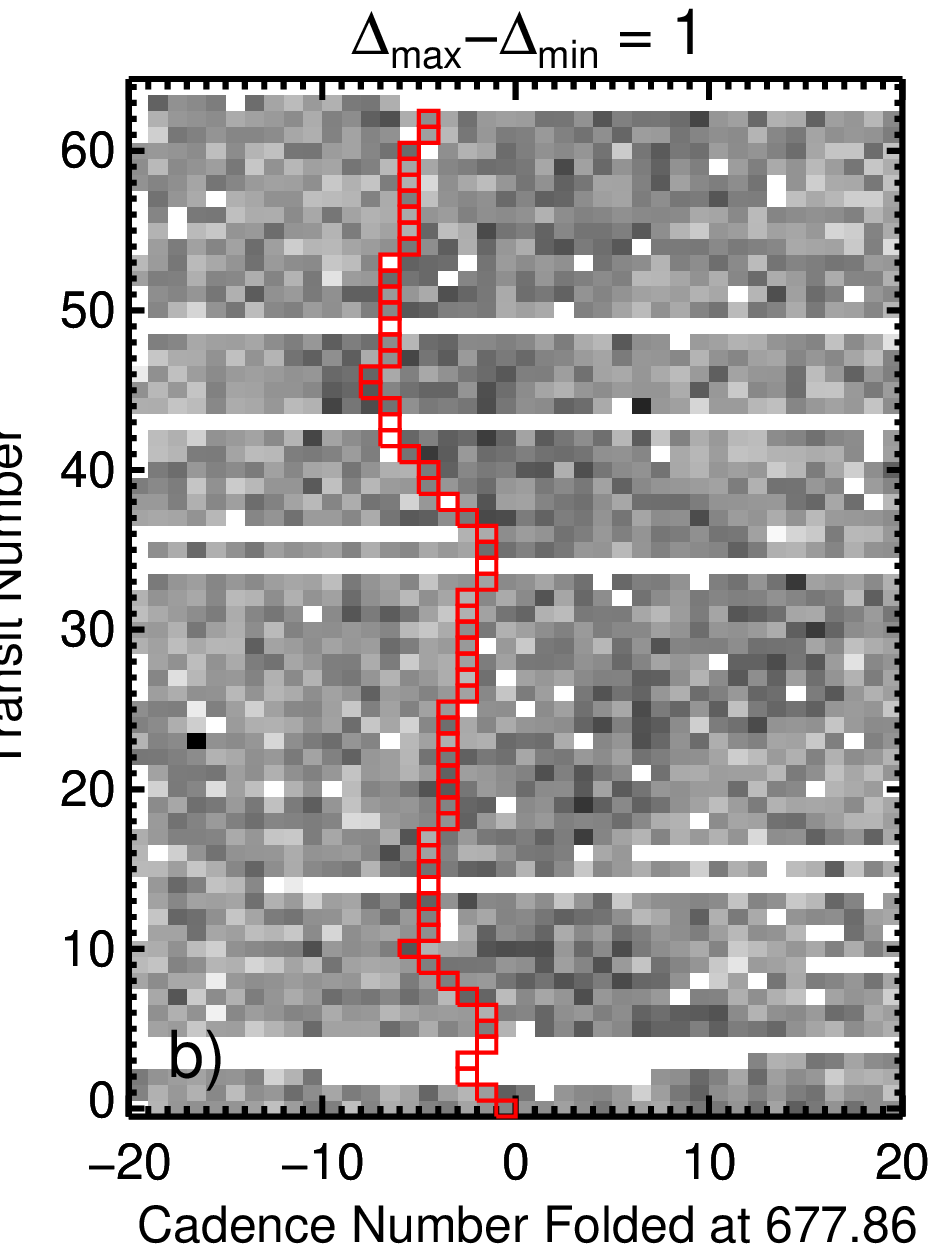}
      \plotone{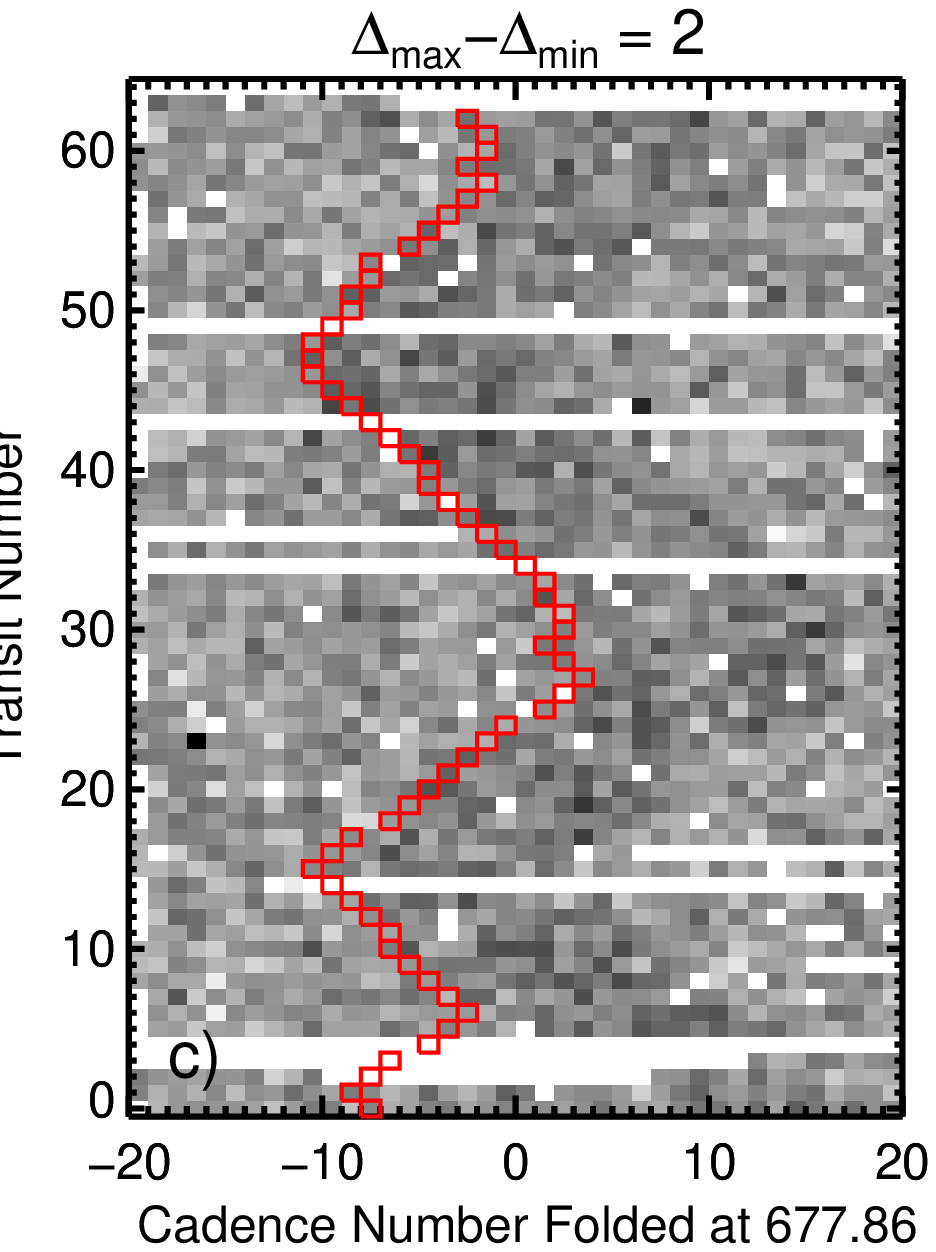}
      \caption{The detection of Kepler-36b. Refer to the caption of Fig.~\ref{fig:ex:crv} for a description of the panels.  Here, we have applied the QATS algorithm to the data shown in the lower panel of Fig.~\ref{fig:ex:lc}, where the transits of planet c have been effectively removed.  The most-likely instants of the transits of planet b are clearly anticorrelated with those of planet c when $\Delta_{\rm max}-\Delta_{\rm min} = 2$.  }
         \label{fig:ex:brv}
\end{figure*}

\section{Summary}

This paper described the Quasiperiodic Automated Transit Search (QATS) algorithm, based upon a specialization of the algorithm suggested by \cite{Kel'Manov522004}.  This algorithm is summarized in pseudocode in Algorithm \ref{alg:qats} with certain definitions provided in \S~\ref{sec:kjsol}. 

 This algorithm extends fixed-period box maximum-likelihood search methods (e.g., \cite{2002A&A...391..369K}) by allowing for a variable interval between transits that is bounded by user-defined minimum and maximum values.  For this reason, the QATS algorithm is well-suited to the unencumbered detection of transiting planets with substantial transit timing variations resulting from N-body interactions in a multi-object planetary system (\cite{2011MNRAS.417L..16G}).  Such transiting planets have proven to be invaluable in characterizing the low mass end of the planetary mass-radius curve (e.g., \cite{2011Natur.470...53L}, \cite{2012Sci...337..556C}).  

The additional computational load of the QATS algorithm relative to that of fixed-period searches is negligible for most searched conditions.   However, in order to produce such an efficient algorithm we require certain preconditions on the data including a uniform cadence without any missing observations.  Such conditions are closely approximated for the {\it Kepler} and {\it CoRoT} missions and may be reached in any circumstance following some guidelines described in \S~\ref{sec:comp}.  

The additional freedom of quasiperiodicity comes at the cost of increased backgrounds. However, we show, in \S~\ref{sec:back}, that the growth of this background is sub-linear with increasing quasiperiodicity and would not likely exceed detection threshold limits in most scenarios. 

The algorithm presented here is generic for any quasiperiodic box-like signal and its application may not be restricted to transiting exoplanets alone.  Going further, altering the box filter to match a different pulse shape could extend the algorithms utility beyond box-shaped pulses; however, this is at the expense of the transit-oriented formalism presented in this paper.  We direct the reader to the more generic work of \cite{Kel'Manov522004} to begin their specialization.'

Implementations of the QATS algorithm may be found at either URL listed below:

 \noindent \begin{verbatim}http://www.cfa.harvard.edu/~jacarter/\end{verbatim}
 
 \noindent or
 
 \noindent \begin{verbatim}http://www.astro.washington.edu/users/agol/\end{verbatim}

\begin{acknowledgements}

The authors are grateful for helpful discussions with A. Kel'Manov, the Kepler TTV/Multibody working group and E. Petigura.  The authors thank the participants of the Sagan Exoplanet Workshop for testing an implementation of the QATS algorithm. J. Carter would like to thank C. Carter, E. Carter and the staff at Phillips House for their inspiration, labor and care during the writing of a portion of this manuscript. J. Carter acknowledges support by NASA through Hubble Fellowship grant HF-51267.01-A awarded by the Space Telescope Science Institute, which is operated by the Association of Universities for Research in Astronomy, Inc., for NASA, under contract NAS 5-26555. E. Agol acknowledges support from NSF Career grant AST 0645416, and thanks Ian Agol for pointing out the usefulness of the Farey sequence in computing the QATS spectrum.

\end{acknowledgements}

\appendix
\section{Derivation of the objective function}

The likelihood that the model $L_n$ represents the data $F_n$ -- which is assumed to be contaminated by additive, Gaussian white noise with characteristic width $\sigma$ -- is
\begin{equation}
	{\cal L}(F_n | L_n; \eta_M, q, \delta) = \left(\sqrt{2 \pi \sigma^2}\right)^{-N} \exp\left[-\frac{1}{2}\sum_{n=0}^{N-1} \left(\frac{F_n-L_n}{\sigma}\right)^2\right].
\end{equation}
We consider the natural logarithm of this likelihood and substitute the definition of $L_n$ from Eqn.~\ref{eq:model},
\begin{eqnarray}
	L(\eta, q, \delta) &\equiv& \log {\cal L}(F_n | L_n; \eta, q, \delta)   \\ \nonumber
		& = & -\frac{N}{2} \log\left(2\pi \sigma^2\right)- \\ \nonumber
		&& \frac{1}{2\sigma^2}\left\{\sum_{n=0}^{N-1} \left( F_n-\sum_{m=1}^M u_{n-n_m}\right)^2\right\}. \label{eq:like}
\end{eqnarray}

For fixed noise (fixed, finite $\sigma$), maximization of the likelihood reduces to the minimization of the final braced expression, which is independent of $\sigma$, in the above log-likelihood.  Expanding this term, 
\begin{eqnarray}
	\sum_{n=0}^{N-1} \left( F_n-\sum_{m=1}^M u_{n-n_m}\right)^2 & = & \sum_{n=0}^{N-1} F_n^2-S^2(\eta,q,\delta) \label{eq:brack}
\end{eqnarray}
where
\begin{eqnarray}
	S^2(\eta, q, \delta) \equiv \sum_{n=0}^{N-1} \left[ 2 \sum_{m=1}^M  F_n u_{n-n_m} - \left(\sum_{m=1}^{M} u_{n-n_m} \right)^2 \right].
\end{eqnarray}
The first term in Eqn.~\ref{eq:brack} is constant with respect to the parameters, so the maximization of the likelihood is reduced to the maximization of $S(\eta,q,\delta)$.  We may write the second term in $S^2(\eta,q,\delta)$ as
\begin{eqnarray}
	\left(\sum_{m=1}^{M} u_{n-n_m} \right)^2 & = & \sum_{m=1}^{M} \sum_{m'=1}^{M} u_{n-n_m} u_{n-n_m'} \\ \nonumber
	 	& = & \sum_{m=1}^{M} \sum_{m'=1}^{M} u_{n-n_m}^2 \delta_{m,m'} \\ \nonumber
		& = & \sum_{m=1}^M u_{n-n_m}^2
\end{eqnarray}
where in the second line we have used the fact that the transits are non-overlapping for $m \neq m'$.  We may then switch the order of summation in $S^2(\eta, q, \delta)$ and let $n-n_m \rightarrow j$ such that
\begin{eqnarray}
	S^2(\eta, q, \delta) &=&  \sum_{m=1}^{M} \left[ 2 \sum_{j=-n_m}^{N-1-n_m}  F_{j+n_m} u_{j} - \sum_{j=-n_m}^{N-1-n_m} u_{j}^2 \right]. \\ \nonumber
\end{eqnarray}
The lower and upper bounds in the sum over $j$ may be replaced with $0$ and $q-1$, respectively, as $u_j$ is non-zero only within those limits (Eqn.~\ref{eq:root}) and Eqns. \ref{eq:cond1}--\ref{eq:cond3} imply that $n_m > 0$ and $N-1-n_m > N-1-n_M > q-1$;
\begin{eqnarray}
	S^2(\eta, q, \delta) &=&  \sum_{m=1}^{M} \left[ 2 \sum_{j=0}^{q-1}  F_{j+n_m} u_{j} - \sum_{j=0}^{q-1} u_{j}^2 \right] \\ \nonumber
\end{eqnarray}
Applying the definition of $u_j$ (Eqn.~\ref{eq:root}) we find that
\begin{eqnarray}
	S^2(\eta, q, \delta) &=&  2 \delta \bar{S}(\eta,q) -M q \delta^2 \\ \nonumber
\end{eqnarray}
where we have isolated the term that depends on the data,
\begin{eqnarray}
	 \bar{S}(\eta,q) &=& \sum_{m=1}^{M} \sum_{j=0}^{q-1}  -F_{j+n_m}.
\end{eqnarray}

The objective function $S(\eta, q, \delta)$ is trivially maximized with respect to the depth of transit.  The depth at maximum likelihood, $\delta_{\rm best}$, obeys the algebraic equation
\begin{eqnarray}
	\left.\frac{\partial S^2(\eta, q, \delta)}{\partial \delta}\right|_{\delta_{\rm best}} & = & 2 \bar{S}(\eta,q) - 2 M q \delta_{\rm best}   =  0
\end{eqnarray}
so that
\begin{eqnarray}
	\delta_{\rm best} & = & \frac{\bar{S}(\eta,q) }{M q}.
\end{eqnarray}
As a result, one need not consider different choices of transit depth in the numerical maximization of $S(\eta, q, \delta)$ with QATS. We only need to consider the marginalized expression over depth,
\begin{eqnarray}
	S^2(\eta, q) & \equiv & S^2(\eta, q, \delta_{\rm best})  =  2 \delta_{\rm best} \bar{S}(\eta,q) -M q \delta_{\rm best}^2\nonumber  \\ 
		& = & 2 \frac{\bar{S}(\eta,q)^2 }{M q} - M q \left(\frac{\bar{S}(\eta,q) }{M q}\right)^2\nonumber  \\ 		& = & \frac{\bar{S}(\eta,q)^2 }{M q}
\end{eqnarray}
and taking a square root, we have our final expression for the object function
\begin{eqnarray}
	S(\eta, q)		& = & \frac{\bar{S}(\eta,q) }{\sqrt{M q}}
\end{eqnarray}

{}

\end{document}